# A Fast Measurement based fixed-point Quantum Search Algorithm

**Ashish Mani and C Patvardhan**

*Department of Electrical Engineering, Faculty of Engineering, Dayalbagh Educational Institute, Dayalbagh, Agra-282110, India*
*ashish.mani@rediffmail.com and cpatvardhan@gmail.com*

**Abstract:** Generic quantum search algorithm searches for target entity in an unsorted database by repeatedly applying canonical Grover's quantum rotation transform to reach near the vicinity of the target entity represented by a basis state in the Hilbert space associated with the qubits. Thus, when qubits are measured, there is a high probability of finding the target entity. However, the number of times quantum rotation transform is to be applied for reaching near the vicinity of the target is a function of the number of target entities present in the unsorted database, which is generally unknown. A wrong estimate of the number of target entities can lead to overshooting or undershooting the targets, thus reducing the success probability. Some proposals have been made to overcome this limitation. These proposals either employ quantum counting to estimate the number of solutions or fixed point schemes. This paper proposes a new scheme for stopping the application of quantum rotation transformation on reaching near the targets by measurement and subsequent processing to estimate the distance of the state vector from the target states. It ensures a success probability, which is at least greater than half for all the ratios of the number of target entities to the total number of entities in a database, which are less than half. The search problem is trivial for remaining possible ratios. The proposed scheme is simpler than quantum counting and more efficient than the known fixed-point schemes. It has same order of computational complexity as canonical Grover`s search algorithm but is slow by a factor of two and requires an additional ancilla qubit.

## 1. Introduction

Quantum Computation has emerged as an exciting field in recent past with development of Shor`s polynomial time Factorizing algorithm [1] and Grover`s Quantum Search Algorithm [2]. Grover`s Quantum Search algorithm finds wide application as a subroutine in solving many types of problems [3]. Grover`s Algorithm is superior to classical search algorithms as it provides for quadratic speed up in search process. It was designed for searching a unique item from an unsorted database containing N items [2]. It is assumed that an oracle exists that will tell whether a selected item satisfies the criterion in one-step. Classically, this problem requires an average of N/2 such oracle calls. Grover's algorithm can find the unique item in $O(\sqrt{N})$ steps, which is much faster than any known classical algorithm. Grover's algorithm was further studied in detail in [4] and tighter bounds were provided for the quantum search. It was observed that the quantum search takes place by rotating the state vector in the Hilbert space associated with the database. If the application of rotation transform is not stopped near the target state, the state vector will either undershoot or overshoot the target state considerably. Further, it has been established that the number of times rotation transform is required to be applied on the state vector is a function of not only the size of the database but also the number of target states. The number of target states is mostly unknown for real world cases, thus, it limits the application of the canonical quantum search algorithm. A number of attempts have been made to solve this problem ranging from employing quantum counting to estimate the number of target states [4] to fixed point quantum searches [5].

In this paper, a novel scheme has been suggested to solve this problem by employing measurement and subsequent processing to estimate the distance of the state vector from the target states. This paper integrates the idea of indirect measurement from instrumentation to estimate the distance from the target state and feedback control to stop near the target state.

The paper is further organized as follows: Section 2 presents existing quantum search algorithms. Section 3 describes and analyzes the proposed algorithm. Conclusion is drawn in Section 4.

## 2. Canonical Quantum Search Algorithms and modifications

The canonical quantum search algorithm proposed by Grover [2] solves a general search problem in which there are N elements that can be represented by n basis states in Hilbert Space i.e. $N \leq 2^n$, where N and n are both positive integers. Let HS = $\{0, 1\}^n$ and let Or : HS → $\{0, 1\}$, where Or represents the oracle, which returns the answer when sampled but no other information is known about Or. Using this framework along with quantum operators, the target state ts, which is a basis state in HS such that Or(ts) = 1, is to be found. It consists of the following steps:

   i. Initializing a set of qubits |s⟩, which represent the solutions and an output qubit.
   ii. Apply quantum rotation gate 'r' times on the qubits to amplify the probability of finding target element.
   iii. Measure qubits |s⟩.
   iv. If the qubits |s⟩ contain the correct information, then stop.
   v. Else, go to step (i).

Grover`s Algorithm is a probabilistic algorithm with computational complexity of O(√N) and its performance has been studied in [4] and was proved to be optimal up to a constant factor. The probability of finding the target element is a function of 'r' given by the following equation [4]:

$$gr(p) = \sin^2((2r+1)\sin^{-1}(\sqrt{p}))  \qquad (1)$$

where p = m/N, m is the number of target states / elements and N is the total number of elements. It is evident that if m is known in advance then the target elements can be easily located. However, m is mostly not known, in advance, in real life problems. Further, if 'r' is selected smaller or even larger than the ideal value, the probability of finding the target elements can reduce considerably. Thus, the probability of finding the solution is reduced not only when less number of quantum rotations are performed but also when more number of quantum rotations are performed than the optimal.

A number of ways have been devised to overcome this limitation of canonical quantum search algorithm, which includes amplitude amplification and quantum counting [4], intelligent heuristic guesses [6-7], Fixed-point algorithms [8] and measurement based fixed-point algorithm [5].

Amplitude amplification and quantum counting techniques increases the requirement of number of different types of quantum operators, which may be practically more difficult to implement, at least initially, and further they will require additional queries [4].

Fixed-point algorithms proposed in [8] makes the quantum search algorithm behave in a similar way as classical search algorithm i.e. Fixed point algorithm does not overshoot the target state and moves monotonically towards it. However, the computational complexity of the proposed algorithm is of O(N) though it is faster by a factor of 2 in comparison to the classical algorithm and is also proven to be asymptotically optimal [5].

Fixed-point quantum algorithms based on measurement have also been suggested in literature [5] which improved on the algorithm proposed in [8]. However, it still has the same computational complexity i.e. O(N) rather than the optimal O(√N) of canonical quantum search algorithm. The fixed-point algorithms were developed for the problems in which expected number of queries is small.

This paper proposes an improved measurement based fixed-point algorithm that provides the same computational complexity as that of canonical quantum search i.e. O(√N) irrespective of the number of target entities by measuring the ancilla qubit and subsequently counting the number of times |0⟩ and |1⟩ have been obtained to compute their ratio. This ratio is used for determining the continuing and stopping criterion of application quantum rotation transform on |s⟩ qubits.

## 3. Proposed Algorithm

The proposed algorithm is illustrated in Fig. 1. It employs |s⟩ qubits whose basis states represent entities in a database. It also includes two ancilla output qubits, |OQ$_1$⟩ and |OQ$_2$⟩. The Grover's canonical quantum rotation gate 'G' and an additional oracle query transformation 'Uf$_1$' along with measurement operators for |s⟩ and |OQ$_2$⟩ have been used. Further, two classical counters, C1 and C0, for counting the number of times |OQ$_2$⟩ collapses to |1⟩ and |0⟩ respectively have been used. The ratio C1/C0 has been employed for estimating the distance of |s⟩ from the target states. The proposed algorithm would be a success if it can be shown that for a specific and unique threshold, Set_Val, of ratio C1/C0, the success probability is at least greater than half for all the ratios of the number of target entities to the total number of entities are less than half in a database. The search problem is trivial for remaining possible ratios of the number of target entities to the total number of entities.

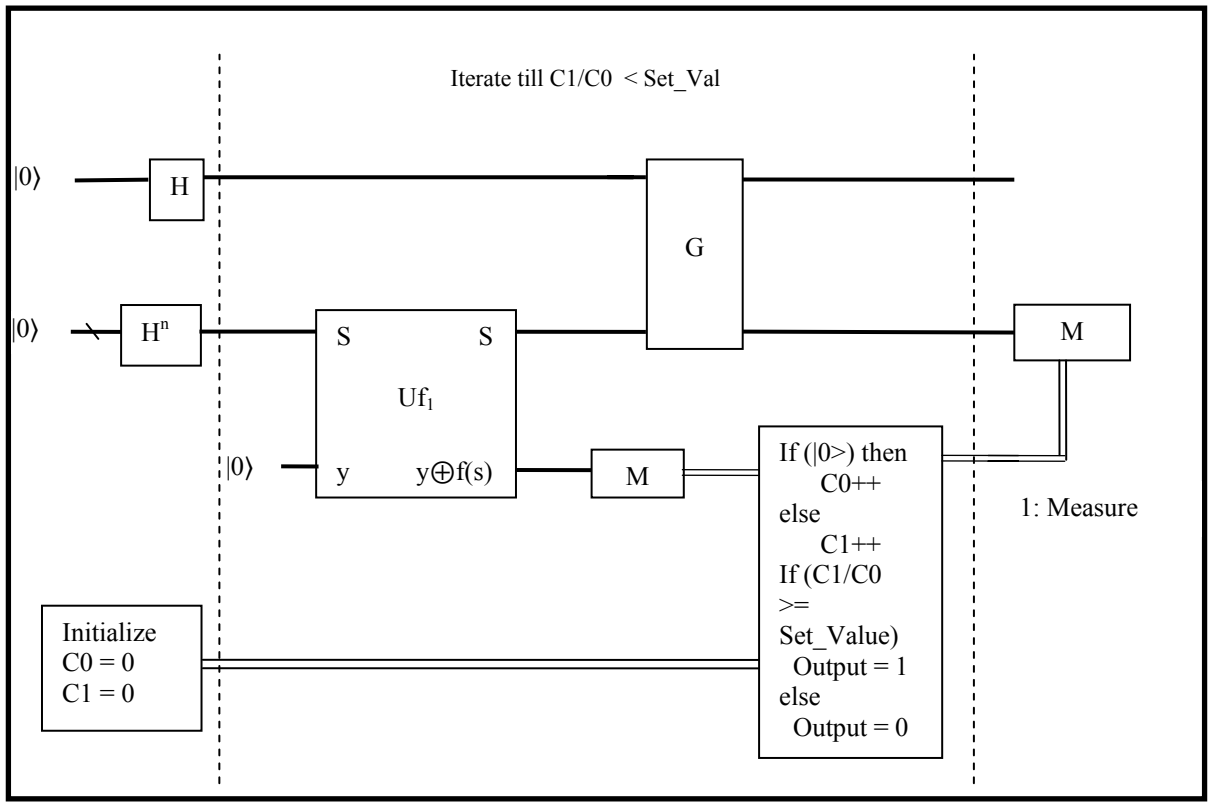

Fig. 1: Quantum circuit model of the proposed Algorithm

The proposed algorithm is as given below:

(i) Initialize counters for measuring the state of the output qubit-2, $|OQ_2\rangle$, post measurement to zero, i.e. C0 for $|0\rangle$ and C1 for $|1\rangle$.

(ii) Initializing a set of qubits $|s\rangle$ and the ancilla qubits, $|OQ_1\rangle$ in state $|0\rangle$.

(iii) Apply Walsh-Hardamard gate on $|s\rangle$ and $|OQ_1\rangle$.

(iv) Set $|OQ_2\rangle$ in state $|0\rangle$.

(v) Apply $Uf_1$ on $|s\rangle$ and $|OQ_2\rangle$.

(vi) Apply G on $|s\rangle$ and $|OQ_1\rangle$ to amplify the probability of finding the target elements.

(vii) Measure the $|OQ_2\rangle$ and increment C0 or C1 depending on the outcome of the measurement.

(viii) If ratio C1/C0 >= Set_Val, then measure the solution qubits, else go to (iv).

(ix) If the qubits contain the correct information, then stop.

(x) Else, go to step (i).

The main modification from canonical Grover's quantum search algorithm is $|OQ_2\rangle$ and its counters, which maintain count of the number of times $|OQ_2\rangle$ upon measurement collapses to $|0\rangle$ i.e. C0 or $|1\rangle$ i.e. C1. The $|OQ_2\rangle$ is in state $|0\rangle$ at the beginning of every iteration and is transformed by $Uf_1$ depending on the distance of $|s\rangle$ from the target state i.e. initially the probability of finding a solution would be low so mostly $|OQ_2\rangle$ would collapse to $|0\rangle$ and as the $|s\rangle$ is transformed in subsequent iterations, the probability of finding the solution would increase, which would result in $|OQ_2\rangle$ collapsing with a high probability in state $|1\rangle$. A relationship can be established between the ratio of C1/C0 and $g_r(p)$ given by equation 1 and some guidelines can be given for choosing Set_Val for measuring the $|s\rangle$ qubits.

## Analysis

The main feature of this algorithm is the establishing a relationship between expected value of C1/C0 (i.e. <C1>/<C0>) and the success probability for all values of p less than ½, which is derived in the following way:

$$\frac{<C1>}{<C0>} = \frac{\sum_{r=0}^{Rn} gr(p)}{\sum_{r=0}^{Rn}(1-gr(p))} \qquad (2)$$

where Rn is the number of rotations at the chosen value of gr(p), which should vary from ½ to 1.

Or

$$\frac{<C1>}{<C0>} = \frac{\int_0^{Rn} gr(p)dr}{\int_0^{Rn}(1-gr(p))dr} \qquad (3)$$

where Rn is the number of rotations at the chosen value of $g_r(p)$, which should vary from ½ to 1.

Therefore,

$$\frac{<C1>}{<C0>} = \frac{(2(sin^{-1}(\sqrt{gr(p)})-sin^{-1}(\sqrt{p}))-(sin(2sin^{-1}(\sqrt{gr(p)}))-sin(2sin^{-1}(\sqrt{p}))))}{(2(sin^{-1}(\sqrt{gr(p)})-sin^{-1}(\sqrt{p}))+(sin(2sin^{-1}(\sqrt{gr(p)}))-sin(2sin^{-1}(\sqrt{p}))))} \qquad (4)$$

Or

$$\frac{<C1>}{<C0>} = \frac{(2X - sin2X - (2\theta - sin2\theta))}{(2X + sin2X - (2\theta + sin2\theta))} \qquad (5)$$

where $X = sin^{-1}(\sqrt{gr(p)})$ and $\theta = sin^{-1}(\sqrt{p})$.

This formula can be employed for setting the ratio of <C1>/<C0> for measuring |s⟩ at any probability of finding the target states provided p is known. However, in most real world examples, p is unknown. In order to overcome this limitation, further analysis has been performed considering the following cases of practical importance to arrive at the rules for assigning values to Set_Val.

Case I:  p = m/N = 1/N i.e. θ ≈ 0 (for large database)

  (a) gr(p) = ½ i.e. X = Π / 4

Inserting value of X and θ in equation (5), <C1>/<C0> = 0.23

  (b) gr(p) = ¾ i.e. X = Π / 3

Inserting value of X and θ in equation (5), <C1>/<C0> = 0.42

  (c) gr(p) = 1 i.e. X = Π / 2

Inserting value of X and θ in equation (5), <C1>/<C0> = 1.00

Case II: p = m/N = ¼ i.e. θ = Π / 6

  (a) gr(p) = ½ i.e. X = Π / 4

Inserting value of X and θ in equation (5), <C1>/<C0> = 0.60

  (b) gr(p) = ¾ i.e. X = Π / 3

Inserting value of X and θ in equation (5), <C1>/<C0> = 1.00

  (c) gr(p) = 1 i.e. X = Π / 2

Inserting value of X and θ in equation (5), <C1>/<C0> = 2.41

Case III: p = m/N = ½ i.e. θ = Π / 4

  (a) gr(p) = ½ i.e. X = Π / 4

Inserting value of X and θ in equation (5), <C1>/<C0> = 1.00

  (b) gr(p) = ¾ i.e. X = Π / 3

Inserting value of X and θ in equation (5), <C1>/<C0> = 1.69

(c) gr(p) = 1 i.e. X = Π / 2

Inserting value of X and θ in equation (5), <C1>/<C0> = 4.50

Therefore, in Case I, the value of <C1>/<C0> should lie between 0.23 and 1.00 for a success probability to lie between ½ and 1.0 if a single target state is present. In Case II, the value of <C1>/<C0> should lie between 0.60 and 2.41 for a success probability to lie between ½ and 1.0 if m = N/4. In Case III, the value of <C1>/<C0> should lie between 1.0 and 4.50 for a success probability to lie between ½ and 1.0 if m = N/2. Thus, if value of **Set_Val is chosen as 1.0** then for all the practical cases (which were represented by Case I to Case III), the probability of success is at least ½ for the boundary Case III (for which a classical randomized algorithm would suffice) or greater than ½ for all other cases. This is illustrated in Table 1.

Table 1: Summary of Calculation for determining a universal value of Set_Val

| Case No. | P = m/N | gr(p) = ½ | gr(p) = ¾ | gr(p) = 1.0 |
|---|---|---|---|---|
| | | <C1>/<C0> | <C1>/<C0> | <C1>/<C0> |
| I | 1/N | 0.23 | 0.42 | **1.00** |
| II | ¼ | 0.60 | **1.00** | 2.41 |
| III | ½ | **1.00** | 1.69 | 4.50 |

The proposed algorithm is successful and asymptotically as fast as the canonical Grover search algorithm, though, it requires an additional query in each iteration and an extra ancilla qubit.

**4. Conclusion**

A fast measurement based fixed- point quantum search algorithm has been proposed which is better than the existing state of art fixed point algorithms. It is asymptotically as fast as the canonical quantum search algorithm and thus optimal up to a constant factor. It improves canonical quantum search algorithms by making it suitable for search problems with multiple target elements. The proposed technique of using feedback control by measuring ancilla qubits will improve other algorithms that currently use Grover`s search as a subroutine like in optimization etc.